\documentclass[journal]{IEEEtran}

\usepackage{url}
\usepackage{ifthen}
\usepackage{cite}
\usepackage[cmex10]{amsmath}
\usepackage[pdftex]{graphicx}
\usepackage{latexsym}
\usepackage{amssymb}
\usepackage{bm}
\usepackage{psfrag}
\usepackage{color}

\newtheorem{definition}{Definition}
\newtheorem{lemma}{Lemma}
\newtheorem{theorem}{Theorem}
\newtheorem{example}{Example}
\newtheorem{remark}{Remark}
\newtheorem{corollary}{Corollary}

\newcommand{\vect}[1]{{\boldsymbol #1}}
\newcommand{\mat}[1]{\mathbf{#1}}

\newcommand{\intg}[1]{\ensuremath{[\![#1]\!]} }
\newcommand{\1}{\mathbb{I}}
\newcommand{\argmax}{\operatornamewithlimits{argmax}}
\newcommand{\qbinom}{\genfrac{[}{]}{0pt}{}}
\newcommand{\hwt}{\mathrm{wt}_{\mathrm{H}}}
\newcommand{\bbZ}{\mathbb{Z}}

\begin{document}
\title{Weight Enumerators and Cardinalities for Number-Theoretic Codes}

\author{Takayuki~Nozaki,~\IEEEmembership{Member,~IEEE}%
  \thanks{This paper has been accepted in IEEE transactions on Information Theory, vol.~68, no.~11, pp.~7165--7173, Nov.\ 2022.}%
  \thanks{The work of T.~Nozaki was supported by Inamori Research Grants and JSPS KAKENHI Grant Number 22K11905.
This paper was presented in part at 2020 the IEEE International Symposium on Information Theory, 2020 \cite{nozaki2020weight}.}%
\thanks{T.~Nozaki is with the Department of Informatics, Yamaguchi University, Yamaguchi 753-8512, JAPAN
    e-mail: tnozaki@yamaguchi-u.ac.jp}
\thanks{Digital Object Identifier 10.1109/TIT.2022.3184776}
}

\maketitle

\begin{abstract}
  The number-theoretic code is a class of codes defined by single or multiple congruences.
  These codes are mainly used for correcting insertion and deletion errors, and for correcting asymmetric errors.
  This paper presents a formula for a generalization of the complete weight enumerator for the number-theoretic codes.
  This formula allows us to derive the weight enumerators and cardinalities for the number-theoretic codes.
  As a special case, this paper provides the Hamming weight enumerators and cardinalities of the non-binary Tenengolts' codes, correcting single insertion or deletion.
  Moreover, we show that the formula deduces the MacWilliams identity for the linear codes over the ring of integers modulo $r$.
\end{abstract}


\IEEEpeerreviewmaketitle

\section{Introduction}
The number-theoretic code \cite{helberg1993coding} is a class of codes defined by single or multiple congruences.
These codes are mainly used for correcting insertion and deletion errors \cite{varshamov1965code,levenshtein1966binary,bibak2018weight,nozaki2019bounded},
and for correcting asymmetric errors \cite{constantin1979constantin-rao,shiozaki1982single}.
In general, the number-theoretic codes are non-linear.

The code rate characterizes the performance of the code and is derived from the number of codewords or cardinality of the code.
The cardinalities of the linear codes are easily derived from the size and rank of the generator or parity-check matrices.
On the other hand, in the case of number-theoretic codes, derivation of the cardinalities is not an easy problem.
Once we can explicitly derive the cardinalities of the number-theoretic codes, we can choose a code with the largest cardinality.
Although it is also non-trivial problem to encode the number-theoretic codes \cite{abdel1998systematic,abroshan2018efficient},
there is a prospect that a low redundant encoding algorithm will be given for the code with the largest cardinality.

The Varshamov-Tenengolts (VT) codes \cite{varshamov1965code} are binary number-theoretic single insertion/deletion correcting codes.
Its cardinality is given by Ginzburg \cite{ginzburg1967certain}, Stanley and Yoder \cite{stanley1972study}.
More precisely, they \cite{ginzburg1967certain}, \cite{stanley1972study} defined an $r$-ary VT code, which is a \textit{natural} generalization of the binary VT codes, and derived the cardinality of this code.
To derive cardinalities of the VT codes, Stanley and Yoder \cite{stanley1972study} derived the Hamming weight enumerators of the codes.
In other words, the Hamming weight enumerators are used for deriving the cardinalities.

Bibak and Milenkovic \cite{bibak2018weight} defined the binary linear congruence (BLC) code, which is a general class of number-theoretic codes, and derived its Hamming weight enumerator.
The BLC code includes the binary codes defined by a single linear congruence, e.g., the binary VT codes \cite{varshamov1965code}, the Levenshtein codes \cite{levenshtein1966binary}, the Helberg codes \cite{helberg1993coding}, and the odd weight codes \cite{nozaki2019bounded}.
Sakurai \cite{sakurai2018explicit} generalized this result, namely, defined the $r$-ary linear congruence code and derived its Hamming weight enumerator.
Moreover, Sakurai \cite{sakurai2018explicit} provided a simple derivation for its Hamming weight enumerator.
However, those classes of codes do not include several useful number-theoretic codes, e.g., the non-binary Tenengolts' codes \cite{tenengolts1984nonbinary}, the shifted VT (SVT) codes \cite{schoeny2017codes}, and the non-binary SVT codes \cite{schoeny2017novel}.

This paper investigates the simultaneous congruences (SC) code, a general class of non-binary codes defined by multiple non-linear congruences.
This paper provides the definition of the SC code, and to my best knowledge, the SC code is first formulated in this paper.
The SC code is a generalization of the $r$-ary linear congruence code and includes the non-binary Tenengolts' codes, the SVT codes, and the non-binary SVT codes.
Furthermore, the SC code includes codes over the ring of integers modulo $r$ and codes defined by the finite Abelian group.
Unfortunately, it is difficult directly to derive the Hamming weight enumerator for the SC codes.
Hence, we define the extended weight enumerator, a generalization of the Hamming weight enumerator, and present a formula for the extended weight enumerator of the SC code.
This formula allows us to derive the weight enumerators and cardinalities for the SC codes.
Moreover, using this result, the paper derives the Hamming weight enumerators and cardinalities of the non-binary Tenengolts' codes.
From this, we clarify the parameters which give maximum cardinality of the non-binary Tenengolts' code.
Furthermore, we show that the formula deduces the MacWilliams identity for the linear codes over the ring of integers modulo $r$.

Summarizing above, the paper contributions are
(i) showing a formula for the extended weight enumerators of the SC codes by generalizing the results in \cite{bibak2018weight,sakurai2018explicit},
(ii) deriving the Hamming weight enumerators and cardinalities of the non-binary Tenengolts' codes,
(iii) clarifying the parameters which give the maximum cardinality of the non-binary Tenengolts' code, and
(iv) deriving the MacWilliams identity for the linear codes over the ring of integers modulo $r$ by the formula for the extended weight enumerator of SC codes.

This paper is an extended version of a conference proceeding \cite{nozaki2020weight}.
This paper includes all the proofs omitted in \cite{nozaki2020weight} and introduces an additional example of SC code, codes defined by the finite Abelian group.
Moreover, we additionally present the fourth contribution above.

The rest of the paper is organized as follows.
Section \ref{sec:pre} introduces the notations and definitions used throughout the paper.
Section \ref{sec:ewe} derives the formula for the extended weight enumerators of the SC codes.
Section \ref{sec:cardrVT} presents the Hamming weight enumerators and cardinalities of the non-binary Tenengolts' codes.
Section \ref{sec:mac} derives the MacWilliams identity for the linear codes over the ring of integers modulo $r$.
Section \ref{sec:conc} concludes the paper.

\section{Preliminaries \label{sec:pre}}
This section gives notations used throughout the paper.
This section also introduces several classes of number-theoretic codes and the weight enumerators. 

\subsection{Notations and Definitions}
Let $\mathbb{Z}$, $\mathbb{Z}^{+}$, and $\mathbb{C}$ be the set of integers, positive integers, and complex numbers, respectively.
Define $\intg{a,b} := \{i\in\mathbb{Z} \mid a \le i \le b\}$ for $a,b\in\mathbb{Z}$, and $\intg{a} := \intg{0,a-1}$ for $a\in\mathbb{Z}^+$.
Let $\mathbb{I}\{P\}$ be the indicator function, which equals $1$ if the proposition $P$ is true and equals $0$ otherwise.
Denote the cardinality of a set $T$, by $|T|$.
Denote a vector of length $n$, by $\langle x_1, x_2,\dots, x_n\rangle$.

For $a,b\in\mathbb{Z}$, we write $a \mid b$ if $a$ divides $b$.
For $a,b\in\mathbb{Z}$ and $n\in\mathbb{Z}^+$, denote $a \equiv b \pmod{n}$ if $(a-b)\mid n$.
Denote the ring of integers modulo $r$, by $\mathbb{Z}_r$.

Let $\mathrm{i}$ be the imaginary unit.
Define $e(x) := \exp (2\pi \mathrm{i} x)$.

\subsection{Number-Theoretic Codes}

Bibak and Milenkovic \cite{bibak2018weight} defined the binary linear congruence (BLC) codes as follows:
\begin{definition}
  \label{def:BLC}
  Suppose $n, m\in\mathbb{Z}^+$, $\vect{h} = \langle h_1,h_2,\dots,h_n \rangle \in\mathbb{Z}^n$, and $a\in\intg{m}$.
  Then, the BLC code of length $n$ with parameters $a, m, \vect{h}$ is defined by
  \begin{align*}
    \mathrm{BLC}_a(n,m,\vect{h}) 
    := 
    \{&  \langle x_1,x_2,\dots,x_n \rangle \in \{0,1\}^n \mid  \\
    &\quad\textstyle \sum_{i=1}^n h_i x_i \equiv a \pmod{m}  \}.
  \end{align*}
\end{definition}

Sakurai \cite{sakurai2018explicit} extended this definition to the $r$-ary case and defined the linear congruence (LC) codes.
\begin{definition}
  \label{def:LC}
  Suppose $n, m\in\mathbb{Z}^+$, $\vect{h} = \langle h_1,h_2,\dots,h_n \rangle \in\mathbb{Z}^n$, and $a\in\intg{m}$.
  Then, the $r$-ary LC code of length $n$ with parameters $a, m, \vect{h}$ is defined by
  \begin{align*}
    \mathrm{LC}_a(n,m,r,\vect{h}) 
    := 
    \{&  \langle x_1,x_2,\dots,x_n \rangle \in \intg{r}^n \mid \\
    &\quad \textstyle \sum_{i=1}^n h_i x_i \equiv a \pmod{m}  \}.
  \end{align*}
\end{definition}

In general, the number-theoretic insertion/deletion correcting codes are defined by multiple non-linear congruences over non-binary elements.
We refer such codes as {\it non-binary multiple non-linear congruence codes} or simply {\it simultaneous congruences (SC) codes}.
\begin{definition} \label{def:SC}
  Suppose $n,r,s\in\mathbb{Z}^+$ and $\vect{m} := \langle m_1,m_2,\dots,m_s \rangle\in (\bbZ^{+})^s$.
  Let $\rho_i: \intg{r}^n \to \mathbb{Z}$ and $a_i \in \intg{m_i}$ for $i\in \intg{1,s}$.
  Define $\vect{\rho} := \langle \rho_1,\rho_2,\dots,\rho_s \rangle$.
  Then, the $r$-ary SC code of length $n$ with parameters $s,\vect{\rho}, \vect{a},\vect{m}$ is defined as
  \begin{align*}
    C_{\vect{\rho},\vect{a},\vect{m}}&(n,r,s) 
    \\
    &:=
    \{ \vect{x}\in\intg{r}^n \mid
    \forall i\in\intg{1,s}~~
    \rho_i(\vect{x}) \equiv a_i \pmod{m_i}
    \}.
  \end{align*}
\end{definition}

\begin{remark}
  For $\vect{h} = \langle h_1,h_2,\dots,h_n \rangle\in\mathbb{Z}^n$,
  define a linear mapping $\ell_{\vect{h}}$ as $\ell_{\vect{h}}(\vect{x}) := \sum_{i=1}^n h_i x_i$.
  Then, we get
  \begin{align*}
    &C_{\ell_{\vect{h}},a,m}(n,2,1)
    =
    \mathrm{BLC}_a(n,m,\vect{h}), \\
    &C_{\ell_{\vect{h}},a,m}(n,r,1)
    =
    \mathrm{LC}_a(n,m,r,\vect{h}).
  \end{align*}
  In words, the SC codes are generalization of the BLC codes and LC codes.
\end{remark}

The non-binary Tenengolts' codes \cite{tenengolts1984nonbinary} are single insertion/deletion correcting codes and defined as follows:
\begin{definition}
  \label{def:rVT}
  Let $n,r\in\mathbb{Z}^+$, $a_1\in\intg{n}$ and $a_2 \in\intg{r}$.
  Define
  \begin{align*}
    &\gamma(\vect{x}) := \sum_{i=1}^{n-1} i \1\{x_i>x_{i+1}\}, &
    &\sigma(\vect{x}) := \sum_{i=1}^{n} x_i .
  \end{align*}
  Then, the $r$-ary Tenengolts' code of length $n$ with parameters $a_1,a_2$ is
  \begin{align*}
    \mathrm{T}_{a_1,a_2}(n,r) 
    &:=
    C_{ \langle \gamma,\sigma \rangle, \langle a_1,a_2 \rangle, \langle n,r \rangle}(n,r,2) \\
    &~=
    \{ \vect{x}\in\intg{r}^n \mid
    \gamma(\vect{x}) \equiv a_1 \pmod{n},
    \\
    &\hspace{23.5mm}\sigma(\vect{x}) \equiv a_2 \pmod{r}  \}.
  \end{align*}

Define
\begin{align*}
  &\gamma_{(\ge)}(\vect{x}) := \sum_{i=1}^{n-1} i \1\{x_i \ge x_{i+1}\},  \\
  &\lambda_{(<)}(\vect{x}) := \sum_{i=1}^{n-1} i \1\{x_i < x_{i+1}\},  \\
  &\lambda_{(\le)}(\vect{x}) := \sum_{i=1}^{n-1} i \1\{x_i \le x_{i+1}\},   
\end{align*}
Then, we have the following variants of the $r$-ary Tenengolts' code.
\begin{align*}
  &\mathrm{T}_{a_1,a_2}^{(\ge)}(n,r)
  :=
  C_{ \langle \gamma_{(\ge)},\sigma \rangle, \langle a_1,a_2 \rangle, \langle n,r \rangle}(n,r,2),
  \\
  &\mathrm{T}_{a_1,a_2}^{(<)}(n,r)
  :=
  C_{ \langle \lambda_{(<)},\sigma \rangle, \langle a_1,a_2 \rangle, \langle n,r \rangle}(n,r,2),
  \\
  &\mathrm{T}_{a_1,a_2}^{(\le)}(n,r)
  :=
  C_{ \langle \lambda_{(\le)},\sigma \rangle, \langle a_1,a_2 \rangle, \langle n,r \rangle}(n,r,2).
\end{align*}
\end{definition}
Note that $\gamma(\vect{x})$, $\gamma_{(\ge)}(\vect{x})$, $\lambda_{(<)}(\vect{x})$, and $\lambda_{(\le)}(\vect{x})$ are non-linear mappings.

\begin{example} \label{ex:VT33}
  Table \ref{tab:VT_33} displays the codewords of $\mathrm{T}_{a_1,a_2}(3,3)$ for $a_1,a_2\in\intg{3}$.
  From this table, we confirm that the non-binary Tenengolts' codes are generally non-linear.
  Note that the cardinalities of the non-binary Tenengolts' codes depend on the parameters $a_1,a_2$.
  We will derive the cardinalities of the non-binary Tenengolts' codes in Section \ref{sec:cardrVT}.
  \begin{table}[t]
    \centering
    \caption{Codewords of non-binary Tenengolts' codes $\mathrm{T}_{a_1,a_2}(3,3)$ \label{tab:VT_33}}
    \begin{tabular}{|c|c|c|c|} \hline
      & $a_2= 0$ & $a_2=1$ & $a_2=2$ \\ \hline
      $a_1=0$ &
      \begin{tabular}{c}
        $\{000, 012, 111,$\\ $210, 222\}$
      \end{tabular}
        & $\{001, 022, 112\}$ & $\{002, 011, 122\}$ \\ \hline
      $a_1=1$ & $\{102, 201 \}$ & $\{100, 202, 211\}$ & $\{101, 200, 212\}$ \\ \hline
      $a_1=2$ & $\{021, 120 \}$ & $\{010, 121, 220\}$ & $\{020, 110, 221\}$ \\ \hline
    \end{tabular}
  \end{table}
\end{example}

\begin{example} \label{ex:VT23}
  Table \ref{tab:VT23} shows the codewords of non-binary Tenengotls' codes $\mathrm{T}_{a_1,a_2}(2,3)$ and
  ones for its variants $\mathrm{T}_{a_1,a_2}^{(\ge)}(2,3)$, $\mathrm{T}_{a_1,a_2}^{(<)}(2,3)$, and $\mathrm{T}_{a_1,a_2}^{(\le)}(2,3)$.
  We will derive the cardinality and property for the variants of Tenengolts' code in Section \ref{ssec:vari_cardqVT}.

  \begin{table}[t]
    \centering
    \caption{Codewords of $\mathrm{T}_{a_1,a_2}(2,3)$, $\mathrm{T}_{a_1,a_2}^{(\ge)}(2,3)$, $\mathrm{T}_{a_1,a_2}^{(<)}(2,3)$, and $\mathrm{T}_{a_1,a_2}^{(\le)}(2,3)$ \label{tab:VT23}}
    \begin{tabular}{|c||c|c|c|c|} \hline
      $\langle a_1, a_2 \rangle$ & $\mathrm{T}_{a_1,a_2}$ & $\mathrm{T}_{a_1,a_2}^{(\ge)}$
      & $\mathrm{T}_{a_1,a_2}^{(<)}$ & $\mathrm{T}_{a_1,a_2}^{(\le)}$  \\ \hline \hline
      $\langle 0,0\rangle$ & $\{00, 12\}$ & $\{12\}$ & $\{00,21\}$ & $\{21\}$ \\ \hline
      $\langle 0,1\rangle$ & $\{01, 22\}$ & $\{01\}$ & $\{10,22\}$ & $\{10\}$ \\ \hline
      $\langle 0,2\rangle$ & $\{02, 11\}$ & $\{02\}$ & $\{11,20\}$ & $\{20\}$ \\ \hline
      $\langle 1,0\rangle$ & $\{21\}$ & $\{00,21\}$ & $\{12\}$ & $\{00,12\}$ \\ \hline
      $\langle 1,1\rangle$ & $\{10\}$ & $\{10,22\}$ & $\{01\}$ & $\{01,22\}$ \\ \hline 
      $\langle 1,2\rangle$ & $\{20\}$ & $\{11,20\}$ & $\{02\}$ & $\{02,11\}$ \\ \hline
    \end{tabular}
  \end{table}
  
\end{example}

\begin{example}
  In this example, we list some codes included in the SC code.
  Define
  \begin{align*}
    \omega(\vect{x})
    &:=
    \ell_{\langle 1,2,\dots,n \rangle}(\vect{x})
    =
    \sum_{i=1}^n i x_i, \\
    \delta(\vect{x})
    &:=
    \sum_{i=1}^{n-1} \1\{x_i > x_{i+1}\}.
  \end{align*}
  For fixed $t, r\in \mathbb{Z}^+$, 
  an integer sequence $\{g_i^{(t, r)}\}_i$ is defined recursively as
  \begin{align*}
    &g_i^{(t, r)} =
    1 + (r-1)\sum_{j=1}^{t} g_{i-j}^{(t, r)}\1\{i-j\ge 1\},
    \qquad \text{for $i\in\mathbb{Z}^+$},
  \end{align*}

  Table \ref{tab:SC} shows some special cases of the SC code.
  Note that the BLC (resp.\ LC) code is also special cases of the SC code with $r=2$ and $s=1$ (resp. $s=1$).

  The codes over $\mathbb{Z}_r$ are also special cases of the SC code.
  In particular, every linear code $\mathcal{L}$ over $\mathbb{Z}_r$ is defined by a full-rank parity check matrix $\mat{H}$ as $\mathcal{L} = \{ \vect{x} \in \mathbb{Z}_r^{n} \mid \mat{H} \vect{x}^T = \vect{0}^T \}$.
  Hence, $\mathcal{L}$ is rewritten as
  \begin{align*}
    \mathcal{L}
    =
    \{ \vect{x}\in\intg{r}^n \mid
    \forall i\in\intg{1,s}~~ \ell_{\vect{h}_i}(\vect{x}) \equiv 0 \pmod{r} \},
  \end{align*}
  where $\vect{h}_i$ represents the $i$-th row of $\mat{H}$.

  The binary codes defined by the finite Abelian group $G$ are also in the SC code as shown in below.
  For any finite Abelian group $G$, there exists $e_1, e_2,\dots, e_s$ such that $G$ is isomorphic to $\mathbb{Z}_{e_1}\times\mathbb{Z}_{e_2}\times \cdots \times \mathbb{Z}_{e_s}$
  and $e_i \mid e_{i+1}$ for $i\in \intg{1,s-1}$.
  Hence, the binary codes defined by $G$ is also defined by $s$ congruences.
  An example of the binary code defined by $G$ is the Constantin-Rao code \cite{constantin1979constantin-rao}.

  \begin{table*}
    \centering
    \caption{Codes included in the SC code \label{tab:SC}}
    \begin{tabular}{|c|c|c|c|c|} \hline
      Code & $r$ & $s$ & $\vect{\rho}$ & $\vect{m}$ \\ \hline \hline
      Binary VT \cite{varshamov1965code} &
      2 & 1 & $\omega$ & $n+1$ \\ \hline
      Levenshtein \cite{levenshtein1966binary} &
      2 & 1 & $\omega$ & $m$ \\ \hline
      Ternary integer \cite{hagiwara2016ordered} &
      3 & 1 & $\ell_{\langle 1,3, \cdots, 2^n-1 \rangle}$ & $2^{n+1}+1$ \\ \hline
      Helberg \cite{helberg1993coding} &
      2 & 1 & $\ell_{\vect{g}^{(t, 2)}}$ & $g_{n+1}^{(t, 2)}$ \\ \hline
      Le-Nguyen \cite{le2016new} &
      $r$ & 1 & $\ell_{\vect{g}^{(t, r)}}$ & $g_{n+1}^{(t, r)}$ \\ \hline
      Odd coefficient \cite{nozaki2019bounded} &
      2 & 1 & $\ell_{\langle 1,3,\cdots,2n-1 \rangle}$ & $2m$ \\ \hline
      AN \cite{shiozaki1982single} &
      2 & 1 & $\ell_{\langle 1,2,\cdots,2^{p-2} \rangle}$ & prime $p$ \\ \hline
      Exponential coefficient \cite{nozaki2019bounded} &
      2 & 1 & $\ell_{\langle 1,2,\cdots,2^{n-1} \rangle}$ & $2^m+1$ \\ \hline \hline

      Shifted VT \cite{schoeny2017codes} &
      2 & 2 & $\langle \omega,\sigma \rangle$ & $\langle m,2 \rangle$ \\ \hline
      Han Vinck-Morita \cite{hanvinck1998codes} &
      2 & 2 & $\langle \omega,\sigma \rangle$ & $\langle n+1,3 \rangle$ \\ \hline
      Non-binary Tenengolts' \cite{tenengolts1984nonbinary} &
      $r$ & 2 & $\langle \gamma, \sigma \rangle$ & $\langle n,r \rangle$ \\ \hline \hline
      Non-binary SVT \cite{schoeny2017novel} & $r$ & 3 &
      $\langle \gamma,\delta,\sigma \rangle$ & $\langle m,2,r \rangle$ \\ \hline \hline

      Linear code over $\mathbb{Z}_r$ & $r$ & $s$ &
      $\langle \ell_{\vect{h}_1}, \ell_{\vect{h}_2}, \dots, \ell_{\vect{h}_s}\rangle$ &
      $\langle r, r, \dots, r\rangle$
      \\ \hline
    \end{tabular}
  \end{table*}
\end{example}

\subsection{Hamming and Extended Weight Enumerator}
Let $\hwt(\vect{x})$ be the Hamming weight for $\vect{x}\in\intg{r}^n$, i.e., $\hwt(\vect{x}) := |\{i \mid x_i \neq 0\}|$.
We define the Hamming weight enumerator for a code $T\subseteq \intg{r}^n$ by 
\begin{equation*}
  \mathcal{H}(T;w)
  =
  \sum_{\vect{x}\in T}w^{\hwt(\vect{x})}.
\end{equation*}

This paper investigates the extended weight enumerator, which is a generalization of the Hamming weight enumerator.
To my best knowledge, it is first defined in the paper.
We will explain how we define the extended weight enumerator in Remark \ref{rem:exp-SC}.
\begin{definition}  \label{def:ex-wt}
  Let $n,r,s\in\mathbb{Z}^+$.
  Let $\rho_i: \intg{r}^n \to \mathbb{Z}$ for $i\in \intg{1,s}$.
  We denote the number of components of $\vect{x}\in\intg{r}^n$ that equal $j$, by $\tau_j(\vect{x})$, i.e., $\tau_j (\vect{x}) := |\{i \mid x_i = j \}|$.
  We define the extended weight enumerator associated to $\vect{\rho} := \langle \rho_1,\rho_2,\dots,\rho_s \rangle$ for a code $T\subseteq \intg{r}^n$ as
  \begin{align*}
    \mathcal{W}(T, \vect{\rho}; \vect{z},\vect{w})
    =
    \sum_{\vect{x}\in T}
    \prod_{i=1}^{s}z_{i}^{\rho_i(\vect{x})} 
    \prod_{j\in \intg{r}} w_{j}^{\tau_j(\vect{x})} .
  \end{align*}
  where $\vect{z} = \langle z_1,z_2,\dots,z_{s} \rangle$ and $\vect{w} = \langle w_0,w_1,\dots,w_{r-1} \rangle$.
\end{definition}

\begin{remark} \label{rem:ele-wt}
  Denote all one vector of length $s$, by $1^s$.
  Define a vector $\vect{w}^{*} := \langle 1,w,w,\dots, w \rangle$ of length $r$.
  Then, the complete weight enumerator $\overline{\mathcal{W}}(T;\vect{w})$ for a code $T$ is 
  \begin{equation*}
    \overline{\mathcal{W}}(T;\vect{w})
    :=
    \mathcal{W}(T, \vect{\rho}; 1^s,\vect{w})
    =
  \sum_{\vect{x}\in T}
  \prod_{j\in\intg{r}} w_{j}^{\tau_j(\vect{x})} .
  \end{equation*}
  Moreover, the Hamming weight enumerator for $T$ is derived from the extended weight enumerator as follows:
  \begin{equation}
    \mathcal{H}(T; w)
    =
    \mathcal{W}(T, \vect{\rho}; 1^s,\vect{w}^*).
    \label{eq:det2ham}
  \end{equation}
  Furthermore, the cardinality of $T$ satisfies
  \begin{equation*}
    |T|
    =
    \mathcal{H}(T; 1)
    =
    \mathcal{W}(T, \vect{\rho}; 1^s,1^r).
  \end{equation*}
\end{remark}

The following lemma is easily derived from the definition of extended weight enumerator.
\begin{lemma}
  Let $T_1,T_2\subseteq \intg{r}^n$.
  If $T_1\cap T_1 = \emptyset$ holds, then
    \begin{equation}
    \mathcal{W}(T_1\cup T_2, \vect{\rho}; \vect{z},\vect{w})
    =
     \mathcal{W}(T_1, \vect{\rho}; \vect{z},\vect{w})
    +\mathcal{W}(T_2, \vect{\rho}; \vect{z},\vect{w}).
    \label{eq:wt-sum}
  \end{equation}
\end{lemma}

\begin{example}
  Similar to Example \ref{ex:VT33}, consider the ternary Tenengolts' code of length 3.
  For each codeword $\vect{x}$ in $\mathrm{T}_{0,0}(3,3)$, Table \ref{tab:ex-t33} summarizes the values of $\gamma(\vect{x}), \sigma(\vect{x})$ and $\tau_i(\vect{x})$ for $i=0,1,2$.
  From this table, the following gives the extended weight enumerator associated to $\langle \gamma,\sigma \rangle$ for $\mathrm{T}_{0,0}(3,3)$:
  \begin{align*}
    &\mathcal{W}(\mathrm{T}_{0,0}(3,3), \langle \gamma,\sigma \rangle; \langle z_1,z_2 \rangle,\vect{w}) \\
    &\quad=
    w_0^3 + z_2^3w_0w_1w_2 + z_2^3w_1^3 + z_1^3z_2^3w_0w_1w_2 + z_2^6w_2^3.
  \end{align*}
  Similarly, the complete and Hamming weight enumerators for $\mathrm{T}_{0,0}(3,3)$ are
  \begin{align*}
    &\overline{\mathcal{W}}(\mathrm{T}_{0,0}(3,3); \vect{w})
     =
     w_0^3 + w_0w_1w_2 + w_1^3 + w_0w_1w_2 + w_2^3, \\
     &\mathcal{H}(\mathrm{T}_{0,0}(3,3); w)
     =
     1 + 2 w^2 + 2 w^3.
  \end{align*}

  \begin{table}[t]
    \centering
    \caption{Codewords in $\mathrm{T}_{0,0}(3,3)$ and values associated to extended weight enumerator \label{tab:ex-t33}}
    \renewcommand{\arraystretch}{1.15}
    \begin{tabular}{|c|c@{~}c@{~~~}c@{~}c@{~}c|c|} \hline
      $\vect{x}$ & $\gamma(\vect{x})$ & $\sigma(\vect{x})$ &
      $\tau_0(\vect{x})$ & $\tau_1(\vect{x})$ & $\tau_2(\vect{x})$ &
      $z_1^{\gamma(\vect{x})}z_2^{\sigma(\vect{x})}\prod_{j=0}^{2}w_j^{\tau_j(\vect{x})}$
       \\ \hline \hline
       000 & 0 & 0 & 3 & 0 & 0 & $w_0^3$ \\ \hline
       012 & 0 & 3 & 1 & 1 & 1 & $z_2^3w_0w_1w_2$ \\ \hline
       111 & 0 & 3 & 0 & 3 & 0 & $z_2^3w_1^3$ \\ \hline
       210 & 3 & 3 & 1 & 1 & 1 & $z_1^3z_2^3w_0w_1w_2$ \\ \hline
       222 & 0 & 6 & 0 & 0 & 3 & $z_2^6w_2^3$ \\ \hline
    \end{tabular}
    \renewcommand{\arraystretch}{1}
  \end{table}
\end{example}

\section{Extended Weight Enumerators for SC Codes \label{sec:ewe}}
This section presents a formula for the extended weight enumerators of the SC codes.
Section \ref{ssec:mr} gives the main results of this section.
Section \ref{ssec:pf-SC} proves them.

\subsection{Main Result and Corollary \label{ssec:mr}}
The following main theorem presents an important property to derive the extended weight enumerators for the SC codes.
\begin{theorem} \label{the:SC-WT}
  Define the SC code (resp.\ extended weight enumerator) as in Definition \ref{def:SC} (resp.\ \ref{def:ex-wt}).
  Denote
  $\textstyle \vect{z}e\bigl(\frac{\vect{u}}{\vect{m}}\bigr)
  := \bigl\langle z_1e\bigl(\frac{u_1}{m_1}\bigr), z_2e\bigl(\frac{u_2}{m_2}\bigr),\dots, z_se\bigl(\frac{u_s}{m_s}\bigr) \bigr\rangle.$
  Then, the following equation holds:
  \begin{align}
    \mathcal{W}&(C_{\vect{\rho},\vect{a},\vect{m}}(n,r,s), \vect{\rho}; \vect{z},\vect{w}) \notag \\
    &=
    \sum_{\vect{u}\in\intg{m_1}\times \intg{m_2} \times \cdots \times \intg{m_s}}
    \mathcal{W}\left(\intg{r}^n, \vect{\rho}; \vect{z}e\left(\tfrac{\vect{u}}{\vect{m}}\right),\vect{w}\right)
    \notag \\
    &\hspace{35mm}\times\prod_{i=1}^{s} \frac{1}{m_i}e \left(- \frac{a_iu_i}{m_i}\right). \notag
  \end{align}
\end{theorem}

\begin{remark}
  Theorem \ref{the:SC-WT} shows that we are able to obtain the extended weight enumerator of the SC code $C_{\vect{\rho},\vect{a},\vect{m}}(n,r,s)$
  if we can derive $\mathcal{W}(\intg{r}^n, \vect{\rho};\vect{z},\vect{w})$.
  Fortunately, in some cases, $\mathcal{W}(\intg{r}^n, \vect{\rho};\vect{z},\vect{w})$ can be written in an explicit formula as shown in the following sections.
\end{remark}

The following corollary gives the Hamming weight enumerators of the BLC and LC codes.
These results coincide {\cite[Thm.~IV.2]{bibak2018weight}} and \cite{sakurai2018explicit}.
\begin{corollary}
  \label{cor:BLC}
  The Hamming weight enumerators of the BLC codes and LC codes are
  \begin{align*}
    &\mathcal{H}(\mathrm{BLC}_{a}(n,m,\vect{h}); w)  \\
      &\qquad=
      \frac{1}{m}\sum_{u\in\intg{m}}
      e\left(-\frac{a u}{m}\right) \prod_{j=1}^{n} \left( 1 + w e\left(\frac{h_j u}{m}\right) \right), \\
    &\mathcal{H}(\mathrm{LC}_{a}(n,m,r,\vect{h}); w)  \\
      &\qquad=
      \frac{1}{m}\sum_{u\in\intg{m}}
      e\left(-\frac{a u}{m}\right) \prod_{j=1}^{n}
      \left( 1 + \sum_{k=1}^{r-1} w e\left(\frac{h_j k u}{m}\right) \right).
  \end{align*}
\end{corollary}

\begin{corollary} \label{cor:sc_cd_card}
  Define
  $\textstyle e\bigl(\frac{\vect{u}}{\vect{m}}\bigr) \allowbreak := \allowbreak \bigl\langle e\bigl(\frac{u_1}{m_1}\bigr), e\bigl(\frac{u_2}{m_2}\bigr),\dots, e\bigl(\frac{u_s}{m_s}\bigr) \bigr\rangle.$
  Consider the SC code given in Definition \ref{def:SC}.
  Then, the complete weight enumerator and the cardinality of the SC code are
  \begin{align}
    &\overline{\mathcal{W}}(C_{\vect{\rho},\vect{a},\vect{m}}(n,r,s); \vect{w}) \notag \\
    &\hspace{5mm}=
    \sum_{\vect{u}\in\intg{m_1}\times \intg{m_2} \times \cdots \times \intg{m_s}}
    \mathcal{W}\left(\intg{r}^n, \vect{\rho}; e\left(\tfrac{\vect{u}}{\vect{m}}\right),\vect{w}\right)  
    \notag \\
    &\hspace{35mm}\prod_{i=1}^{s} \frac{1}{m_i}e \left(- \frac{a_iu_i}{m_i}\right). \notag
    \\
    &|C_{\vect{\rho},\vect{a},\vect{m}}(n,r,s)| \notag \\
    &\hspace{5mm}=
    \sum_{\vect{u}\in\intg{m_1}\times \intg{m_2} \times \cdots \times \intg{m_s}}
    \mathcal{W}\left(\intg{r}^n, \vect{\rho}; e\left(\tfrac{\vect{u}}{\vect{m}}\right),1^r\right)  
    \notag \\
    &\hspace{35mm}\prod_{i=1}^{s} \frac{1}{m_i}e \left(- \frac{a_iu_i}{m_i}\right). \notag
  \end{align}
\end{corollary}

  Corollary \ref{cor:sc_cd_card} shows that the extended weight enumerator for $\intg{r}$ is also used for deriving the complete/Hamming weight enumerator of an SC code.
  On the other hand, when we derive the cardinality of an SC code,
  Corollary \ref{cor:sc_cd_card} only uses $\mathcal{W}(\intg{r}, \vect{\rho}; \vect{z}, 1^r) = \sum_{\vect{x}\in \intg{r}^n} \prod_{i=1}^s z_i^{\rho_i(\vect{x})}$.

  From Corollary \ref{cor:sc_cd_card}, when we have an explicit formula of $\mathcal{W}(\intg{r}, \vect{\rho}; \vect{z}, 1^r)$,
  we can obtain the cardinality of an SC code in $\mathcal{O}(\prod_{i=1}^{s} m_i)$.
  Because $m_i$ is a polynomial of $n$ in most SC codes, the complexity of the derivation is also a polynomial of $n$.
  It is much smaller than the complexity $\mathcal{O}(r^n)$ of enumerating all codewords of the SC code.

\subsection{Proof of Main Result and Corollary \label{ssec:pf-SC}}
The following well-known identity is used in the proof.
\begin{lemma} \label{lem:indi}
  For any $A\in\mathbb{Z}, m\in\mathbb{Z}^+$, the following holds:
  \begin{equation*}
    \1\{A\equiv 0 \pmod{m}\}
    =
    \1\{m \mid A\}
    =
    \frac{1}{m}\sum_{j\in\intg{m}}
    e\left(\frac{Aj}{m} \right).
  \end{equation*}
\end{lemma}
The following lemma gives the key technique of the proof.
The technique is well-known in the graph-based codes (e.g., see \cite{frey1997factor}).
\begin{lemma} \label{lem:cms}
  The code membership function
  $\1 \{\vect{x}\in C_{\vect{\rho},\vect{a},\vect{m}}(n,r,s)\}$ is factorized as follows:
  \begin{align*}
    &\1\{\vect{x}\in C_{\vect{\rho},\vect{a},\vect{m}}(n,r,s)\}
    \\
    &\hspace{5mm}=
    \prod_{i=1}^{s} \1 \{ \rho_i(\vect{x}) - a_i \equiv 0 \pmod{m_i} \}
  \end{align*}
\end{lemma}
By combining those lemmas, the code membership function is written as in the following lemma.
\begin{lemma} \label{lem:cms-2}
  The code membership function
  $\1 \{\vect{x}\in C_{\vect{\rho},\vect{a},\vect{m}}(n,r,s)\}$ is written as
  \begin{align*}
    & \1\{\vect{x}\in C_{\vect{\rho},\vect{a},\vect{m}}(n,r,s)\}
    \notag \\
    &\hspace{5mm}=
  \prod_{i=1}^{s} \sum_{u_i\in\intg{m_i}} \frac{1}{m_i}
  e\biggl( -\frac{a_iu_i}{m_i} \biggr)
  \left( e\biggl( \frac{ u_i}{m_i} \biggr) \right)^{\rho_i(\vect{x})}.
  \end{align*}
\end{lemma}
\begin{IEEEproof}
Note that $e(x+y) = e(x)e(y)$ and $e(xy) = \{e(x)\}^y$.
Definition \ref{def:SC} and Lemmas \ref{lem:indi} and \ref{lem:cms} lead
\begin{align*}
  \1\{\vect{x}&\in C_{\vect{\rho},\vect{a},\vect{m}}(n,r,s)\}
  \notag \\
  &=
  \prod_{i=1}^{s} \1 \{ \rho_i(\vect{x}) - a_i \equiv 0 \pmod{m_i} \}
  \notag \\  &=  
  \prod_{i=1}^{s} \sum_{u_i\in\intg{m_i}} \frac{1}{m_i}
  e\left( \frac{(\rho_i(\vect{x}) - a_i)u_i}{m_i} \right)
  \notag \\  &=  
  \prod_{i=1}^{s} \sum_{u_i\in\intg{m_i}} \frac{1}{m_i}
  e\left( - \frac{a_iu_i}{m_i} \right)
  \left( e\left( \frac{ u_i}{m_i} \right) \right)^{\rho_i(\vect{x})}.
\end{align*}
\end{IEEEproof}

\subsubsection{Proof of Theorem \ref{the:SC-WT}}
By combining Lemma \ref{lem:cms-2} and Definition \ref{def:ex-wt}, we obtain the theorem as follows:
\begin{align*}
  \mathcal{W}&(C_{\vect{\rho},\vect{a},\vect{m}}(n,r,s), \vect{\rho}; \vect{z},\vect{w}) 
   \\  &=
   \sum_{\vect{x}\in \intg{r}^n}
   \prod_{i=1}^{s}z_{i}^{\rho_i(\vect{x})} 
   \prod_{j\in\intg{r}} w_{j}^{\tau_j(\vect{x})}
   \1 \{\vect{x}\in C_{\vect{\rho},\vect{a},\vect{m}}(n,r,s)\}
   \\  &=
   \sum_{\vect{x}\in \intg{r}^n}
   \prod_{j\in\intg{r}} w_{j}^{\tau_j(\vect{x})}  
   \prod_{i=1}^{s} \\
   &\hspace{20mm}\sum_{u_i\in\intg{m_i}}\frac{1}{m_i}
  e\left( - \frac{a_iu_i}{m_i} \right)
  \left( e\left( \frac{ u_i}{m_i} \right) z_i \right)^{\rho_i(\vect{x})}
   \\  &=
    \sum_{\vect{u}\in\intg{m_1}\times \intg{m_2} \times \cdots \times \intg{m_s}}
   \left( \prod_{i=1}^{s} \frac{1}{m_i}  e\left( - \frac{a_iu_i}{m_i} \right)
   \right) \\
   &\hspace{5mm}\times
   \sum_{\vect{x}\in \intg{r}^n}
   \prod_{j\in\intg{r}} w_{j}^{\tau_j(\vect{x})}
   \left\{
   \prod_{i=1}^{s}\left( e\left( \frac{ u_i}{m_i} \right) z_i \right)^{\rho_i(\vect{x})}
   \right\}
   \\  &=
    \sum_{\vect{u}\in\intg{m_1}\times \intg{m_2} \times \cdots \times \intg{m_s}}
    \prod_{i=1}^{s} \frac{1}{m_i}  e\left( -\frac{a_iu_i}{m_i} \right)
    \\
   &\hspace{30mm}\times\mathcal{W}(\intg{r}^n, \vect{\rho}; \vect{z}e(\vect{u}/\vect{m}),\vect{w}).
\end{align*}
\hfill\IEEEQED

\subsubsection{Proof of Corollary \ref{cor:BLC}}
The following identity holds
\begin{equation*}
  \mathcal{W}(\intg{r}^n, \ell_{\vect{h}}; z, \vect{w})
  =
  \prod_{j=1}^{n} \sum_{k\in\intg{r}} w_k z^{h_j k}.
\end{equation*}
This equation and Theorem \ref{the:SC-WT} lead
\begin{align*}
  \mathcal{W}&(\mathrm{LC}_{a}(n,m,r,\vect{h}), \ell_{\vect{h}}; z, \vect{w})
  \\
  &=
  \sum_{u\in\intg{m}}
   \frac{1}{m}e \left(- \frac{au}{m}\right)
   \prod_{j=1}^{n}
   \sum_{k\in \intg{r}} w_k z^{h_j k} e\left( \frac{h_j k u}{m} \right).
\end{align*}
By combining this equation and Remark \ref{rem:ele-wt}, we get the Hamming weight enumerator for the LC codes:
\begin{align*}
  \mathcal{H}&(\mathrm{LC}_{a}(n,m,r,\vect{h}); w)
  \\ &=
  \mathcal{W}(\mathrm{LC}_{a}(n,m,r,\vect{h}), \ell_{\vect{h}}; 1, \vect{w}^{*})
  \\ &=
  \sum_{u\in\intg{m}}
   \frac{1}{m}e \left(- \frac{au}{m}\right)
  \prod_{j=1}^{n}\left(1 + w \sum_{k=1}^{r-1} e\left(\frac{h_j ku}{m}\right) \right) .
\end{align*}
Since $\mathrm{BLC}_a(n,m,\vect{h}) = \mathrm{LC}_a(n,m,2,\vect{h})$, we also get the Hamming weight enumerator for the BLC codes by substituting $r=2$ into the equation above.
\hfill\IEEEQED

\begin{remark} \label{rem:exp-SC}
  Now, we explain how we define the extended weight enumerator based on the proof of Theorem \ref{the:SC-WT}.
  The code membership function $\1 \{\vect{x}\in C_{\vect{\rho},\vect{a},\vect{m}}(n,r,s)\}$ generates the factor $\prod_{i=1}^s e(u_i/m)^{\rho_i(\vect{x})}$.
  The extended weight enumerator is defined by adding parameters $\vect{z}$ to the complete weight enumerator for including this factor.
\end{remark}

\section{Cardinalities and Hamming Weight Enumerators for non-binary Tenengolts' Codes \label{sec:cardrVT}}
This section derives the Hamming weight enumerators and cardinalities for non-binary Tenengolts' codes by using Theorem \ref{the:SC-WT}.
Moreover, we show the parameters which give maximum cardinality of the $r$-ary Tenengolts' code of length $n$.
Section \ref{ssec:nota4} gives the notations used in this section.
Section \ref{ssec:main_rVT} presents the results above and Section \ref{ssec:prf_rVT} proves them.
Section \ref{ssec:ex_rVT} shows a numerical example.

\subsection{Notations Used in This Section \label{ssec:nota4}}
For $a,b\in\mathbb{Z}$, let $(a,b)$ be the greatest common divisor of $a$ and $b$.
For $n\in\mathbb{Z}^{+}$, let $\phi(n)$ and $\mu(n)$ be Euler's totient function and M\"{o}bius function, respectively, i.e., when $n$ has a prime factorization $n= p_1^{i_1}p_2^{i_2}\cdots p_k^{i_k}$,
\begin{align*}
  \phi (n)
  &=
  n \prod_{j=1}^{k} \biggl(1 - \frac{1}{p_j} \biggr) ,
  \\  %
  \mu (n)
  &=
  \begin{cases}
    (-1)^{k}, & \text{if~~} i_1 = i_2 = \cdots = i_k = 1, \\
    0, & \text{otherwise}.
  \end{cases}
\end{align*}

The value $\eta_{d}\in\mathbb{C}$ is called a primitive $d$-th root of unity if $\eta_{d}^{i}\neq 1$ for $i\in \intg{1,d-1}$ and $\eta_d^d = 1$ hold.
For $d\in \mathbb{Z}^+$ and $a\in \mathbb{Z}$, denote Ramanujan's sum, by $c_d(a)$, i.e.,
\begin{equation}
  c_d (a)
  :=
  \sum_{j\in \intg{1,d}, (j,d) = 1} e \left( \frac{aj}{d} \right)
  =
  \phi(d) \frac{ \mu(d/(a,d)) }{ \phi(d/(a,d)) }.
  \label{eq:Ramanu}
\end{equation}
Notice that $c_d(a) = c_d(b)$ if $a\equiv b \pmod{d}$.

For non-negative integers $a,b, t_0,t_1,\dots,t_{r-1}$, denote $\vect{t} :=\langle t_0, t_1, \dots, t_{r-1}\rangle$ and we define the binomial coefficient and multinomial coefficient as
\begin{align*}
  \binom{a+b}{a} &:= \frac{(a+b)!}{a! b!}, \\
  \binom{\sum_{i=0}^{r-1} t_i }{\vect{t}}
  &:=
  \binom{\sum_{i=0}^{r-1} t_i }{t_0, t_1, \dots, t_{r-1}}
  =
  \frac{(\sum_{i=0}^{r-1} t_i)! }{\prod_{i=0}^{r-1}t_i !}.
  \\  
\end{align*}

For $n\in\mathbb{Z}^+$, the $q$-integer is defined by 
\begin{equation*}
  [n]_q := 1 + q + q^2 + \cdots + q^{n-1}.
\end{equation*}
Similarly, for non-negative integers $a,b, t_0,t_1,\dots,t_{r-1}$, we define the $q$-factorial, the $q$-binomial coefficient, and the $q$-multinomial coefficient as follows
\begin{align*}
  [a]!_q
  &:= \textstyle \prod_{i=1}^{a} [i]_q
  = [a]_q[a-1]_q \cdots [1]_q,
  \\ 
  \qbinom{a+b}{a}_q
  &:=
  \frac{[a+b]!_{q}}{[a]!_q[b]!_q},
  \\ 
  \qbinom{\sum_{i=0}^{r-1}t_{i}}{\vect{t}}_q
  &:=
  \qbinom{\sum_{i=0}^{r-1}t_{i}}{t_0,t_1,\dots,t_{r-1}}_q
  =
  \frac{[\sum_{i=0}^{r-1}t_{i}]!_{q}}{\prod_{i=0}^{r-1}[t_{i}]!_{q}},
\end{align*}
where $\vect{t} = \langle t_0,t_1,\dots, t_{r-1} \rangle$.
Hereafter, we drop the subscript $q$ if it is clear from the context.

\subsection{Main Result \label{ssec:main_rVT}}
The following theorem presents the Hamming weight enumerators and cardinalities of the non-binary Tenengolts' codes.
\begin{theorem} \label{the:rVT_card}
  Define the non-binary Tenengolts' code as in Definition \ref{def:rVT}.
  For any $a_1 \in \intg{n}, a_2\in \intg{r}$, the following equations give the Hamming weight enumerator and cardinality of the $r$-ary Tenengolts' code of length $n$ with parameters $a_1, a_2$:
  \begin{align*}
    &\mathcal{H}(\mathrm{T}_{a_1,a_2}(n,r); w)
     =    
    \frac{1}{nr}
    \sum_{ d\in\mathbb{Z}^+, d\mid n }
    \sum_{ e\in\mathbb{Z}^+, e\mid r }
    c_d(a_1) c_{e}(a_2) 
    \\
    &\hspace{30mm}\times \left\{ 1 - w^{d} + r w^{d} \1 \{ e \mid d \} \right\}^{\frac{n}{d}},
    \\ 
    &|\mathrm{T}_{a_1,a_2}(n,r)|
    =
    \frac{1}{nr} 
    \sum_{d=1}^{n}  c_{d}(a_1)  r^{\frac{n}{d}} (r,d)
    \1 \{ d \mid n \} \1 \{ (r,d) \mid a_2 \}.
  \end{align*}
\end{theorem}
The following corollary shows that the non-binary Tenengolts' code with parameters $a_1=a_2=0$ has the largest cardinality for any $n$ and $r$.
\begin{corollary} \label{cor:rVT}
  Let $n,r\in\bbZ^+$, $a_1\in \intg{n}$ and $a_2\in \intg{r}$.
  For any $n,r,a_1,a_2$, the following holds
  \begin{equation*}
    |\mathrm{T}_{a_1,a_2}(n,r)|
    \le 
    |\mathrm{T}_{0,0}(n,r)|
    =
    \frac{1}{nr} 
    \sum_{d\in \bbZ^+, d \mid n}  \phi(d)  r^{\frac{n}{d}} (r,d) .
  \end{equation*}
\end{corollary}

\subsection{Proofs \label{ssec:prf_rVT}}
The following two lemmas are used to lead Theorem \ref{the:rVT_card}.
The following lemma is provided by MacMahon \cite[Ch.~VI]{macmahon1915combinatory}.
\begin{lemma} \label{lem:mac}
  Suppose $\vect{t} := \langle t_0,t_1,\dots, t_{r-1} \rangle\in\intg{n}^r$ satisfies $n = \sum_{i=0}^{r-1}t_i$.
  Define $S(\vect{t}) := \{\vect{x}\in \intg{r}^n \mid \forall i\in\intg{r}~~\tau_i(\vect{x}) = t_i  \}$.
  Then, the following identity holds:
  \begin{equation*}
    \sum_{\vect{x}\in S(\vect{t})} q^{\gamma(\vect{x})}
      =
      \qbinom{n}{\vect{t}}_q
  \end{equation*}
\end{lemma}
The following lemma is shown by Hagiwara and Kong \cite[Lem.~2.4]{hagiwara2020applications}.
\begin{lemma} \label{lem:hag}
  Let $\eta_d \in \mathbb{C}$ be a primitive $d$-th root of unity.
  Let $a,b$ be non-negative integers.
  Assume $d \mid (a+b)$.
  Then, we get
  \begin{align*}
    \lim_{q\to \eta_d} \qbinom{a + b}{a}_q
    =
    \binom{\frac{a + b}{d}}{\frac{a}{d}}
    \1\{d \mid a\}\1\{d \mid b \}.
  \end{align*}
\end{lemma}

This lemma is easily generalized as the following.
\begin{lemma} \label{lem:qbinom_prim}
  Let $t_0,t_1,\dots,t_{r-1}$ be non-negative integers.
  Let $\eta_d \in \mathbb{C}$ be a primitive $d$-th root of unity.
  Assume $d \mid \sum_{i=0}^{r-1} t_i$.
  Then, we get 
  \begin{equation*}
    \lim_{q\to \eta_d}
    \qbinom{\sum_{j\in\intg{r}}t_i}{\vect{t}}_q
    =
    \binom{\sum_{j\in\intg{r}}t_i/d}{\vect{t}/d}
    \prod_{i\in\intg{r}} \1\{ d \mid t_i \},
  \end{equation*}
  where $\vect{t} = \langle t_0,t_1,\dots,t_{r-1} \rangle$.
\end{lemma}
\begin{IEEEproof}
  Firstly, we suppose $d \mid t_i$ for all $i\in\intg{r}$.
  The $q$-multinomial coefficient is factorized as follows:
  \begin{align*}
    \qbinom{\sum_{i=0}^{r-1}t_i}{\vect{t}}_q
    &=
    \qbinom{\sum_{j=0}^{r-1} t_j}{t_{r-1}}_q
    \qbinom{\sum_{j=0}^{r-2} t_j}{t_{r-2}}_q
    \cdots
    \qbinom{\sum_{j=0}^{1} t_j}{t_{1}}_q
    \\ &=
    \prod_{k=1}^{r-1} \qbinom{\sum_{j=0}^k t_j}{t_k}_q.
  \end{align*}
  Note that $d \mid \sum_{j=0}^k t_j$ holds for all $k\in \intg{1,r-1}$.
  By applying Lemma \ref{lem:hag} to this equation, we get
  \begin{equation*}
    \lim_{q\to \eta_d} \qbinom{\sum_{i=0}^{r-1}t_i}{\vect{t}}_q
    =
    \prod_{k=1}^{r-1} \binom{\sum_{j=0}^k t_j/d}{t_k/d}
    =
    \binom{\sum_{j=0}^{r-1} t_j/d}{\vect{t}/d}.
  \end{equation*}

  Next, we suppose that there exists $i\in\intg{r}$ such that $d \nmid t_i$.
  Since $t_i$ is a non-negative integer, it is expressed as $t_i = a_i d + b_i$, where $b_i \in\intg{d}$.
  From the above, there exists $i$ such that $b_i \neq 0$.
  Since $d \mid \sum_{i=0}^{r-1} t_i$ holds, we get $d \mid \sum_{i=0}^{r-1} b_i$.
  Hence, we have $\sum_{i=0}^{r-1} b_i \ge d$.
  Note that $q$-multinomial coefficient is rewritten as
  \begin{equation}
    \qbinom{\sum_{i=0}^{r-1}t_i}{\vect{t}}_q
    =
    \qbinom{\sum_{i=0}^{r-1}a_i d}{ d \vect{a} }_q
    \frac{\prod_{j=1}^{\sum_{i=0}^{r-1} b_i} [\sum_{i=0}^{r-1}a_i d +j]_q }
         {\prod_{i=0}^{r-1} \prod_{j=1}^{b_i} [a_id+j]_q},
         \label{eq:pr-lm-mul-hag1}
  \end{equation}
  where $\vect{a} = \langle a_0, a_1, \dots, a_{r-1} \rangle$.
  For all non-negative integer $a$ and $j\in \intg{d}$, the following holds
  \begin{equation*}
    \lim_{q\to \eta_d} [a d + j]_{q}
    =
    \begin{cases}
      0, & \text{if~} j \equiv 0 \pmod{d}, \\
      \sum_{k=0}^{j-1} \eta_{d}^k \neq 0, & \text{otherwise}.
    \end{cases}
  \end{equation*}
  From this, the denominator of the second factor in \eqref{eq:pr-lm-mul-hag1} does not contain zero factor.
  On the other hand, the numerator of the second factor in \eqref{eq:pr-lm-mul-hag1} contain zero factor, since $\sum_{i=0}^{r-1} b_i \ge d$.
  Hence, $\lim_{q\to\eta_d} \qbinom{\sum_{i=0}^{r-1}t_i}{\vect{t}}_q = 0$.
\end{IEEEproof}

\subsubsection{Proof of Theorem \ref{the:rVT_card}}

We divide the proof into several lemmas.
\begin{lemma}  \label{lem:marge} 
  Let $n,r\in\bbZ^+$.
  Define $T_n := \{ \langle t_0,t_1,\dots,t_{r-1} \rangle \in\allowbreak\intg{n}^r \mid\allowbreak \sum_{i\in\intg{r}}t_i = n\}$.
  The following holds:
  \begin{align*}
    \mathcal{W}(\intg{r}^n, \langle \gamma,\sigma \rangle ; \langle q,z \rangle,\vect{w})
    =
    \sum_{\vect{t}\in T_n}
    \qbinom{n}{\vect{t}}
    \prod_{j\in\intg{r}} w_j^{t_j} z^{j t_j}.
  \end{align*}
\end{lemma}
\begin{IEEEproof}
  Lemma \ref{lem:mac} gives
  \begin{align*}
    \mathcal{W}(S(\vect{t}), \langle \gamma,\sigma \rangle; \langle q,z \rangle,\vect{w})
    =
    \qbinom{n}{\vect{t}}  \prod_{j\in\intg{r}} w_j^{t_j} z^{j t_j}.
  \end{align*}
  For $\vect{t},\vect{t}' \in T_n$ ($\vect{t}\neq \vect{t}'$), $S(\vect{t})\cap S(\vect{t}') = \emptyset$ holds.
  Moreover, $\bigcup_{\vect{t}\in T_n} S(\vect{t}) = \intg{r}^n$.
  Equation \eqref{eq:wt-sum} yields
  \begin{align*}
  \mathcal{W}(\intg{r}^n, \langle \gamma,\sigma \rangle; \langle q,z \rangle,\vect{w})
   &=
  \sum_{\vect{t}\in T_n}  \mathcal{W}(S(\vect{t}), \langle \gamma,\sigma \rangle; \langle q,z \rangle,\vect{w})
  \\ &=
  \sum_{\vect{t}\in T_n} \qbinom{n}{\vect{t}}
  \prod_{j\in\intg{r}} w_j^{t_j} z^{j t_j}.
  \end{align*}
  This concludes the proof.
\end{IEEEproof}
\begin{lemma}
  For any $u_1\in\intg{n}, u_2\in\intg{r}$, we get
  \begin{align}
    \mathcal{W}&
    \left(\intg{r}^n, \langle \gamma,\sigma \rangle; \left\langle e\left(\tfrac{u_1}{n}\right),e\left(\tfrac{u_2}{r}\right) \right\rangle ,\vect{w}\right)
    \notag \\
    &=
    \left( \sum_{i\in\intg{r}} w_i^{\frac{n}{(n,u_1)}} e\left(\frac{i n u_2}{(n,u_1) r}\right) \right)^{(n,u_1)}.
    \label{eq:total_comp}
  \end{align}
  In particular, for $\vect{w} = \vect{w}^{*} = \langle 1,w,w,\dots,w \rangle$, we have
  \begin{align}
    \mathcal{W}&
    \left(\intg{r}^n, \langle \gamma,\sigma \rangle; \left\langle e\left(\tfrac{u_1}{n}\right),e\left(\tfrac{u_2}{r}\right) \right\rangle,\vect{w}^*\right)
    \notag \\  &=
    \left(1 - w^{\frac{n}{(n,u_1)}} + rw^{\frac{n}{(n,u_1)}} \1\{ r \mid \tfrac{n u_2}{(n,u_1)} \} \right)^{(n,u_1)}.
    \label{eq:total_ham}
  \end{align}
\end{lemma}
\begin{IEEEproof}
  For a fixed $u_1$, define $d:= \frac{n}{(n,u_1)}$.
  Then, $e(u_1/n)$ is a primitive $d$-th root of unity.
  Combining Lemmas \ref{lem:qbinom_prim} and \ref{lem:marge} gives
  \begin{align*}
  \mathcal{W}&(\intg{r}^n, \langle \gamma,\sigma \rangle; \langle e(\tfrac{u_1}{n}),z \rangle,\vect{w} )
   \\ &=
  \sum_{\vect{t}\in T_n}  \binom{n/d}{\vect{t}/d}
  \prod_{j\in\intg{r}} w_j^{t_j} z^{j t_j}  \1\{ d \mid t_j \}
  \\ &=
  \sum_{\vect{t}' \in T_{n/d}}  \binom{n/d}{\vect{t}'}
  \prod_{j\in\intg{r}} (w_j z^j)^{d t_j'}
  \\ &=
  \textstyle
  \left( \sum_{i\in\intg{r}} w_i^d z^{id} \right)^{n/d},
  \end{align*}
  where we replace $t_i = d t_i'$ in the second equality.
  Substituting $z=e(u_2/r)$ in this equation leads \eqref{eq:total_comp}.

  Lemma \ref{lem:indi} shows
  \begin{equation*}
    \textstyle
    \sum_{j=1}^{m-1}  e\left(\frac{Aj}{m} \right)
    =
    m \1\{m \mid A\} - 1.
  \end{equation*}
  Combining this identity and \eqref{eq:total_comp} gives \eqref{eq:total_ham}.
\end{IEEEproof}

\noindent\qquad\textit{Proof of Theorem \ref{the:rVT_card}:}
The Hamming weight enumerator is derived as follows:
\begin{align*}
  \mathcal{H}&(\mathrm{T}_{a_1,a_2}(n,r); w)
  \\ &=
  \mathcal{W}(\mathrm{T}_{a_1,a_2}(n,r), \langle \gamma,\sigma \rangle; (1,1),\vect{w}^*)
  \\ &=
  \frac{1}{nr} \sum_{u_1\in\intg{n}} \sum_{u_2\in\intg{r}}
  e\Bigl(-\frac{a_1u_1}{n}\Bigr)e\Bigl(-\frac{a_2u_2}{r}\Bigr)
  \\  &\hspace{8mm}\times
  \mathcal{W} \left(\intg{r}^n, \langle \gamma,\sigma \rangle; \left(e\left(\tfrac{u_1}{n}\right),e\left(\tfrac{u_2}{r}\right)\right),\vect{w}^*\right)
  \\ &=
  \frac{1}{nr}   
  \sum_{\substack{d\in\bbZ^+, \\ d\mid n}}
  \sum_{\substack{u_1\in\intg{n}, \\ (u_1,n) = d}}
  \sum_{\substack{e\in\bbZ^+, \\ e\mid r}}
  \sum_{\substack{u_2\in\intg{r}, \\ (u_2,r) = e}}
  e\Bigl(-\frac{a_1u_1}{n}\Bigr)e\Bigl(-\frac{a_2u_2}{r}\Bigr) \\
  &\hspace{8mm}\times\left(1 - w^{n/d} + rw^{n/d} \1\{ r \mid \tfrac{n}{d} u_2\} \right)^{d}
  \\ &=
  \frac{1}{nr} 
  \sum_{\substack{d\in\bbZ^+, \\ d\mid n}}
  \sum_{\substack{u_1'\in\intg{n/d}, \\ (u_1',n/d) = 1}}
  \sum_{\substack{e\in\bbZ^+, \\ e\mid r}}
  \sum_{\substack{u_2'\in\intg{r/e}, \\ (u_2',r/e) = 1}} e\biggl(-\frac{a_1u_1'}{n/d}\biggr)
  \\  &\hspace{8mm}\times
  e\biggl(-\frac{a_2u_2'}{r/e}\biggr) \left(1 - w^{n/d} + rw^{n/d} \1\{ \tfrac{r}{e} \mid \tfrac{n}{d} \} \right)^{d}
  \\ &=
  \frac{1}{nr} 
  \sum_{d'\in\bbZ^+, d' \mid n}  c_{d'}(a_1)
  \\
  &\hspace{15mm}\sum_{e'\in\bbZ^+,e' \mid r}  \left(1 - w^{d'} + rw^{d'} \1\{ e' \mid d' \} \right)^{\frac{n}{d'}}
  c_{e'}(a_2).
\end{align*}
Here, the first, second, and third equality follow from \eqref{eq:det2ham}, Theorem \ref{the:SC-WT}, and \eqref{eq:total_ham}, respectively;
The fourth equality is derived by replacing $u_1 = d u_1', u_2 = e u_2'$ and using $[(a,u_2')=1]\land[a\mid b u_2'] \Rightarrow [a \mid b]$;
The fifth equality is derived by replacing $d' = n/d, e'= r/e$ and using \eqref{eq:Ramanu}.

Note that the following identity holds for $a\in\bbZ^+, b\in\bbZ$
\begin{equation*}
  \sum_{d\in\bbZ^+, d\mid a}c_{d}(b)
  =
  \sum_{j \in\intg{a}} e\biggl(\frac{jb}{a}\biggr)
  =
  a \1 \{ a \mid b \}.
\end{equation*}
By substituting $w=1$ in the Hamming weight enumerator, we get the cardinality as follows:
\begin{align*}
  |\mathrm{T}_{a_1,a_2}&(n,r)|
  \\ &=
  \frac{1}{nr} 
  \sum_{d\in\bbZ^+, d \mid n}  c_{d}(a_1) 
  \sum_{e\in\bbZ^+, e \mid r}  r^{\frac{n}{d}} \1\{ e \mid d \} 
  c_{e}(a_2)
  \\ &=
  \frac{1}{nr} 
  \sum_{d\in\bbZ^+, d \mid n}  c_{d}(a_1) r^{\frac{n}{d}}
  \sum_{e\in\bbZ^+}   \1\{ e \mid (d,r) \} c_{e}(a_2)
  \\ &=
  \frac{1}{nr} 
  \sum_{d\in\bbZ^+, d \mid n}  c_{d}(a_1) r^{\frac{n}{d}} (d,r) \1 \{ (d,r) \mid a_2 \}.
\end{align*}
This concludes the proof.
\hfill\IEEEQED

\subsubsection{Proof of Corollary \ref{cor:rVT}}
For any $a,b\in\mathbb{Z}^+$, we get $1 = \1\{a \mid 0\} \ge \1\{a \mid b\}$.
For any $d\in\mathbb{Z}^+$, $a\in \mathbb{Z}$, $c_d(0) \ge c_d(a)$ holds.
Hence, we get Corollary \ref{cor:rVT}.
\hfill\IEEEQED

\subsection{Numerical Example \label{ssec:ex_rVT}}
Consider the case of $n=3,r=3$.
For any $a_1,a_2\in\intg{3}$, Theorem \ref{the:rVT_card} shows
\begin{align*}
  |\mathrm{T}_{a_1,a_2}(3,3)|
  &=
  \textstyle
  \frac{1}{9}
  \sum_{d\mid 3} 3^{3/d} c_{d}(a_1)  (3,d) \1 \{ (3,d) \mid a_2 \}
  \\ &=
  3 c_1 (a_1) + \1 \{3 \mid a_2 \} c_3(a_1)
\end{align*}
Note that $c_1(0) = 1, c_3(0) = 2, c_3(1) = c_3(2) = -1$.
We get
\begin{align*}
  |\mathrm{T}_{a_1,a_2}(3,3)|
  &=
  \begin{cases}
    5, & \text{if }a_1 = 0 \text{ and } a_2 = 0, \\
    2, & \text{if }a_1 = 1,2 \text{ and } a_2 = 0, \\ 
    3, & \text{if } a_2 = 1,2.
  \end{cases}
\end{align*}
This result is confirmed by Table \ref{tab:VT_33}.

\subsection{Property for Variants of Non-binary Tenengolts' code \label{ssec:vari_cardqVT}}

For $\vect{x}=\langle x_1,x_2,\dots, x_n \rangle$, denote the vector reversing the order of $\vect{x}$, by $\bar{\vect{x}}$, i.e.,
$\bar{\vect{x}}:=\langle x_{n}, \dots, x_2, x_1 \rangle$.
The following theorem shows that the variants of the non-binary Tenengolts' code are essentially equivalent.
\begin{theorem} \label{thm:equivalence}
  Let $n, r\in \mathbb{Z}^+$.
  For $a_1\in\intg{n}$, define $\bar{a}_1 := (n-a_1)\mathbb{I}\{a_1 \neq 0\}$ and
  \begin{align*}
    a_1' :=
    \begin{cases}
     (n-a_1)\mathbb{I}\{a_1 \neq 0\}, & \text{odd $n$}, \\
     n/2 - a_1 + n\mathbb{I}\{a_1 > n/2\}, & \text{even $n$}. \\
    \end{cases}
  \end{align*}
  Then, for any $a_1 \in \intg{n}$ and $a_2 \in \intg{r}$.
  \begin{align}
    &\mathrm{T}_{a_1, a_2}^{(<)}(n,r) = \{ \bar{\vect{x}} \mid \vect{x}\in \mathrm{T}_{\bar{a}_1, a_2}(n,r) \},
    \label{eq:equiv1}  \\
    &\mathrm{T}_{a_1, a_2}^{(\le)}(n,r) = \{ \bar{\vect{x}} \mid \vect{x}\in \mathrm{T}_{\bar{a}_1, a_2}^{(\ge)}(n,r) \},
    \label{eq:equiv2} \\
    &\mathrm{T}^{(\le)}_{a_1, a_2}(n,r) = \mathrm{T}_{a'_1, a_2}(n,r),
    \label{eq:equiv3} \\
    &\mathrm{T}^{(\ge)}_{a_1, a_2}(n,r) = \mathrm{T}^{(<)}_{a'_1, a_2}(n,r) .
    \label{eq:equiv4} 
  \end{align}
\end{theorem}
\begin{IEEEproof}
Obviously, $\sigma(\bar{\vect{x}}) = \sigma(\vect{x})$ for any $\vect{x}\in\intg{r}^n$.
Moreover, for any $\vect{x}\in\intg{r}^n$, we get 
\begin{align*}
  \lambda_{(<)}(\bar{\vect{x}})
  &=
  n(n-1) - \gamma(\vect{x}).
\end{align*}
Hence, $\lambda_{(<)}(\bar{\vect{x}})\equiv a_1 \pmod{n}$ if and only if $\gamma(\vect{x}) \equiv  \bar{a}_1 \pmod{n}$.
Thus, we obtain \eqref{eq:equiv1}.
Similarly, we get \eqref{eq:equiv2}.

For any $\vect{x}\in\intg{r}^{n}$, we have
\begin{align*}
  \gamma(\vect{x}) + \lambda_{(\le)}(\vect{x})
   =
   \frac{n(n-1)}{2}.
\end{align*}
Hence, we get
\begin{align*}
  \gamma(\vect{x}) + \lambda_{(\le)}(\vect{x})
  \equiv
  \begin{cases}
    0  \pmod{n}, & \text{odd $n$}, \\
    \frac{n}{2} \pmod{n}, & \text{even $n$}.
  \end{cases}
\end{align*}
Hence, $\lambda_{(\le)}(\vect{x}) \equiv a_1 \pmod{n}$ if and only if $\gamma(\vect{x}) \equiv a'_1 \pmod{n}$.
Thus, we obtain \eqref{eq:equiv3}.
Similarly, we get \eqref{eq:equiv4}.
\end{IEEEproof}

Theorem \ref{thm:equivalence} leads the cardinalities and Hamming weight enumerators for the variants of non-binary Tenengolts' codes. 
\begin{theorem} \label{thm:vari_card}
  Let $n,r \in \mathbb{Z}^+$.
  For $a_1\in\intg{n}$, define $\bar{a}_1$ and $a'_1$ as in Theorem \ref{thm:equivalence}.
  Define
  \begin{equation*}
    \bar{a}'_1
    :=
    \begin{cases}
      a_1, & \text{odd $n$}, \\
      n/2 + a_1 - n\mathbb{I}\{a_1 \ge n/2 \}, & \text{even $n$}.
    \end{cases}
  \end{equation*}
  Then, for any $a_1\in\intg{n}$ and $a_2\in\intg{r}$,
  \begin{align}
    &\mathcal{H}(\mathrm{T}_{a_1,a_2}^{(<)}(n,r);z) = \mathcal{H}(\mathrm{T}_{\bar{a}_1,a_2}(n,r);z),  \label{eq:hwd_1}\\
    &\mathcal{H}(\mathrm{T}_{a_1,a_2}^{(\le)}(n,r);z) = \mathcal{H}(\mathrm{T}_{a'_1,a_2}(n,r);z),  \label{eq:hwd_2}\\
    &\mathcal{H}(\mathrm{T}_{a_1,a_2}^{(\ge)}(n,r);z) = \mathcal{H}(\mathrm{T}_{\bar{a}'_1,a_2}(n,r);z), \label{eq:hwd_3} \\
    &|\mathrm{T}_{a_1,a_2}^{(<)}(n,r)| = |\mathrm{T}_{\bar{a}_1,a_2}(n,r)|, \label{eq:card_1} \\
    &|\mathrm{T}_{a_1,a_2}^{(\le)}(n,r)| = |\mathrm{T}_{a'_1,a_2}(n,r)|, \label{eq:card_2} \\
    &|\mathrm{T}_{a_1,a_2}^{(\ge)}(n,r)| = |\mathrm{T}_{\bar{a}'_1,a_2}(n,r)|. \label{eq:card_3}
  \end{align}
\end{theorem}
\begin{IEEEproof}
We get \eqref{eq:hwd_1} and \eqref{eq:equiv3} from \eqref{eq:equiv1} and \eqref{eq:hwd_2}, respectively.
Equation \eqref{eq:equiv4} leads $\mathcal{H}(\mathrm{T}_{a_1,a_2}^{(\ge)}(n,r);z) = \mathcal{H}(\mathrm{T}_{a'_1,a_2}^{(<)}(n,r);z)$.
Combining this and \eqref{eq:hwd_2}, we get \eqref{eq:hwd_3}.
By substituting $z=1$ into \eqref{eq:hwd_1}, \eqref{eq:hwd_2}, and \eqref{eq:hwd_3}, we obtain \eqref{eq:card_1}, \eqref{eq:card_2}, and \eqref{eq:card_3}, respectively.
\end{IEEEproof}

The following corollary gives the parameters $\langle a_1, a_2 \rangle$ which attain the maximum cardinality.
\begin{corollary} \label{cor:att-car}
  For any $n, r\in \mathbb{Z}^+$, the following holds
  \begin{align*}
    &\argmax_{\langle a_1, a_2 \rangle} |\mathrm{T}_{a_1,a_2}^{(<)}(n,r)|
    \ni \langle 0, 0 \rangle,
    \\ 
    &\argmax_{\langle a_1, a_2 \rangle} |\mathrm{T}_{a_1,a_2}^{(\ge)}(n,r)|
    \ni
    \begin{cases}
      \langle 0, 0 \rangle, & \text{odd $n$}, \\
      \langle n/2, 0 \rangle, & \text{even $n$}, \\      
    \end{cases}
    \\ 
    &\argmax_{\langle a_1, a_2 \rangle} |\mathrm{T}_{a_1,a_2}^{(\le)}(n,r)|
    \ni
    \begin{cases}
      \langle 0, 0 \rangle, & \text{odd $n$}, \\
      \langle n/2, 0 \rangle, & \text{even $n$}. \\      
    \end{cases}
  \end{align*}
\end{corollary}
\begin{IEEEproof}
  Combining Corollary \ref{cor:rVT} and Theorem \ref{thm:vari_card}, we obtain the corollary.
\end{IEEEproof}

  As shown in Example \ref{ex:VT23} and Corollary \ref{cor:att-car}, when $n$ is even,
  the variants $\mathrm{T}_{a_1,a_2}^{(\ge)}(n,r), \mathrm{T}_{a_1,a_2}^{(\le)}(n,r)$ of the non-binary Tenengolts' code do not attain the maximum cardinality for $a_1 = a_2 = 0$.

\section{MacWilliams Identity for Complete Weight Enumerators of Linear Codes \label{sec:mac}}

Recall that $\mathcal{L}$ is a linear code over $\mathbb{Z}_r$.
Let $\mathcal{L}_{\bot}$ be the dual code of $\mathcal{L}$, i.e.,
\begin{equation*}
  \mathcal{L}_{\bot}
  :=
  \{ \vect{y}\in \mathbb{Z}_r \mid
  \vect{y}\vect{x}^T = 0 ~~\forall \vect{x}\in\mathcal{L}\}.
\end{equation*}
Note that $\mathcal{L}_{\bot}$ is represented by a full-rank parity-check matrix $\mat{H}$ for $\mathcal{L}$, as
\begin{equation*}
  \mathcal{L}_{\bot}
  =
  \{ \vect{u} \mat{H} \mid \vect{u} \in \mathbb{Z}_r^{s}  \}.
\end{equation*}

  The MacWilliams identity \cite{macwilliams1977theory} gives the complete weight enumerator of a code from one of its dual code.
  This identity is widely investigated in the coding theory, e.g., \cite{shi2019onadditive,shi2015note}.
Now, we will derive the MacWilliams identity for the complete weight enumerator of the linear code over $\mathbb{Z}_r$ (e.g., see \cite{wood2008lecture}) from Theorem \ref{the:SC-WT}.
\begin{corollary}
  Let $\mathcal{L}$ be the linear code over $\mathbb{Z}_r$, and $\mathcal{L}_{\bot}$ its dual code.
  For a fixed $\vect{w} = \langle w_0, w_1, \dots, w_{r-1}\rangle$,
  define $v_i := \sum_{k=0}^{r-1} w_k e(ik/r)$ for $i\in \intg{r}$.
  Then, the following identity holds:
  \begin{equation*}
    \overline{\mathcal{W}}(\mathcal{L}; \vect{w})
    =
    \frac{1}{r^s} \overline{\mathcal{W}}(\mathcal{L}_{\bot}; \vect{v}).
  \end{equation*}
\end{corollary}
\begin{IEEEproof}
  Denote $\vect{h}_i = \langle h_{i,1}, h_{i,2}, \dots, h_{i,n} \rangle$.
  Note that
  \begin{equation*}
    \mathcal{W}(\intg{r}^n, \langle \ell_{\vect{h}_1}, \ell_{\vect{h}_2}, \dots, \ell_{\vect{h}_s}\rangle; \vect{z}, \vect{w} )
    =
    \prod_{j=1}^n \sum_{k=0}^{r-1} w_k \prod_{i=1}^s z_i^{h_{i,j} k}.
  \end{equation*}
  Denote $e(\vect{u}/r) = \langle e(u_1/r), e(u_2/r), \dots, e(u_s/r) \rangle$.
  From Theorem \ref{the:SC-WT}, we get
  \begin{align*}
    \overline{\mathcal{W}}( \mathcal{L}; \vect{w} )
    &=
    \frac{1}{r^s} \sum_{\vect{u} \in \intg{r}^{s}}
    \mathcal{W}( \intg{r}^n, \langle \ell_{\vect{h}_1}, \ell_{\vect{h}_2}, \dots, \ell_{\vect{h}_s}\rangle; e(\vect{u}/r), \vect{w})
    \\ &=
    \frac{1}{r^s} \sum_{\vect{u} \in \intg{r}^{s}}
    \prod_{j=1}^n \sum_{k=0}^{r-1} w_k  e\left(\frac{k \sum_{i=1}^s h_{i,j} u_i}{r}\right)
    \\ &=
    \frac{1}{r^s} \sum_{\vect{y} \in \mathcal{L}_{\bot}}
    \prod_{j=1}^n \sum_{k=0}^{r-1} w_k  e\left(\frac{k y_j}{r}\right)
    \\ &=
    \frac{1}{r^s} \sum_{\vect{y} \in \mathcal{L}_{\bot}}
    \prod_{i=0}^{r-1}\left(\sum_{k=0}^{r-1} w_k  e\left(\frac{i k}{r}\right) \right)^{\tau_{i}(\vect{y})}.
  \end{align*}
  Recall that $\overline{\mathcal{W}}(\mathcal{L}_{\bot}; \vect{w}) = \sum_{\vect{y}\in \mathcal{L}_{\bot}} \prod_{i=0}^{r-1}w_i^{\tau_{i}(\vect{y})}$.
  Combining these, we complete the proof.
\end{IEEEproof}

\section{Conclusion \label{sec:conc}}
This paper has derived a formula for the extended weight enumerator of the SC code.
As a special case, this paper has also provided the Hamming weight enumerators and cardinalities of the non-binary Tenengolts' codes.
Moreover, we have shown that the formula deduces the MacWilliams identity for the complete weight enumerator of the linear codes over $\mathbb{Z}_r$.

\section*{Acknowledgments}
The author wishes to thank to Dr.\ Manabu Hagiwara (Chiba Univ., Japan) for his valuable advice.
The author also would like to thank the anonymous reviewers for their careful reading and their insightful comments and suggestions.

\ifCLASSOPTIONcaptionsoff
  \newpage
\fi

\end{document}